\documentstyle[aps,prc,epsfig]{revtex}

\begin{document}
\draft
\title{Transverse momentum fluctuations 
due to temperature variation \\
in high-energy nuclear collisions}
\author{R. Korus$^1$, St. Mr\'owczy\'nski$^{1,2}$, 
M. Rybczy\'nski$^1$, and Z. W\l odarczyk$^1$\\[3mm]}

\address{$^1$Institute of Physics, \'Swi\c etokrzyska Academy,
ul. Konopnickiej 15, PL - 25-406 Kielce, Poland \\[2mm]
$^2$So\l tan Institute for Nuclear Studies, 
ul. Ho\.za 69, PL - 00-681 Warsaw, Poland \\[2mm]}

\date{20-th August 2001}

\maketitle

\begin{abstract}

The event-by-event $p_{\perp}-$fluctuations due to the temperature variations 
are considered. The so-called $\Phi-$measure is computed and shown to be a 
linear function of the temperature variance. The fluctuations of temperature
naturally explain the data on $\Phi(p_{\perp})$ in proton-proton and central 
Pb-Pb collisions but independent measurements of the temperature fluctuations
are needed to confirm the explanation. Feasibility of such an event-by-event 
measurement is discussed. 

\end{abstract}

\vspace{0.5cm}

 

\section{Introduction}

Predictions of different models of heavy-ion collisions are often 
quite similar when averaged characteristics of the collisions are considered. 
Fluctuations are usually much more sensitive to the collision dynamics and
consequently can be helpful in discriminating among the models. 
Since large acceptance detectors, which have recently become common, 
make possible a detailed analysis of individual collisions the study of the
event-by-event fluctuations appears to be a very promising field of 
high-energy heavy-ion physics, see \cite{Hei00} for a review.

The $p_{\perp}-$fluctuations in proton-proton and central Pb-Pb 
collisions at 158 GeV per nucleon have been recently measured \cite{App99}
on event-by-event basis. To eliminate trivial `geometrical' fluctuations due 
to the impact parameter variation, the so-called $\Phi-$measure \cite{Gaz92} 
has been used. $\Phi$ is constructed is such a way that it is exactly the same 
for nucleon-nucleon (N-N) and nucleus-nucleus (A-A) collisions if the A-A 
collision is a simple superposition of the N-N interactions. In that case 
$\Phi$ is independent of the centrality of an A-A collision. Moreover,
$\Phi$ equals zero when the inter-particle correlations are entirely 
absent. The critical analysis of the $\Phi-$measure can be found in 
\cite{Tra00,Uty01}. In the central Pb-Pb collisions the measured value of 
$\Phi(p_{\perp})$ equals $4.6 \pm 1.5$ MeV while the preliminary result for 
proton-proton interactions in the same acceptance region is $5 \pm 1$ MeV 
\cite{App99}. Although the two values are close, the mechanisms 
behind them seem to be different. It has been shown \cite{App99} that 
the correlations, which have short range in the momentum space, like
those due to the quantum statistics, are responsible for the positive 
value of $\Phi(p_{\perp})$ in the central Pb-Pb collisions. When these 
correlations are subtracted $\Phi(p_{\perp})=0.6 \pm 1.0$ MeV \cite{App99}. 
Our calculations have also demonstrated \cite{Mro98,Mro99} that the effect 
of Bose statistics of pions reduced by the hadron resonances fully explains 
the experimentally observed $\Phi(p_{\perp})=4.6 \pm 1.5$ in the central 
Pb-Pb collisions. In the p-p case, the situation seems to be opposite - 
the short range correlations provide a negligible contribution to 
$\Phi(p_{\perp})$ while the whole effect is due to the long range 
fluctuations. Thus, the data suggest that the dynamical long range 
correlations are reduced in the central Pb-Pb collisions (when compared 
to p-p) with the short range caused by the Bose statistics being amplified. 
The former feature is a natural consequence of the system's evolution towards 
the thermodynamic equilibrium. The amplification of the quantum statistics 
effect results from the increased particle population in the final state 
phase-space. Since various dynamical correlations contribute to 
$\Phi(p_{\perp})$ the question arises what is the dynamical correlation 
in the nucleon-nucleon interactions which appears to be absent in the 
central nucleus-nucleus collisions.

In the recent paper of the two of us \cite{Kor01}, the correlation, which 
couples the average $p_{\perp}$ to the event multiplicity $N$, has been 
studied. The correlation is convincingly evidenced in the p-p collisions 
\cite{Aiv88}. The approximate analytical formula of $\Phi(p_{\perp})$ 
as a function of the correlation strength has been derived and then 
the numerical simulation has been performed taking into account the finite 
detector's acceptance. The effect of the correlation 
$\langle p_{\perp} \rangle$ {\it vs.} $N$ has been shown to be very weak 
if the particles from a small acceptance region are studied. Consequently, 
the correlation is far too small to explain the preliminary experimental 
value of  $\Phi (p_{\perp})$ in the proton-proton collisions \cite{App99}. 

Our aim here is to discuss another possible mechanism responsible for the 
finite value of  $\Phi(p_{\perp})$ in p-p interactions \cite{App99}.
Namely, we analyze the effect of the temperature fluctuations. Its role 
in shaping the particle spectra has been studied before \cite{Uty01a}. 
Here, the temperature, or more generally the slope parameter of the 
$p_{\perp}-$distribution, is assumed to vary from event to event. 
We compute $\Phi(p_{\perp})$ and find it to be a linear function 
of the temperature variance. As is well known \cite{Lan80}, the $T-$variance 
is directly related to the system's heat capacity. The idea to exploit 
the relationship to determine the specific heat of matter produced in nuclear 
collisions has been formulated in \cite{Sto95,Shu98}, see also \cite{Ste99}. 
Since the temperature 
fluctuations have been shown \cite{Wil00} to yield the so-called 
non-extensive Tsallis statistics \cite{Tsa88} (with a power-law instead 
of an exponential energy distribution) we express $\Phi(p_{\perp})$ 
by the non-extensivity parameter $q$. Further, we perform the 
numerical simulation of the p-p interactions with the effect of detector's 
finite acceptance taken into account. The temperature fluctuations are 
shown to explain in a natural way the data on $\Phi(p_{\perp})$ in 
proton-proton and central Pb-Pb collisions. Finally, we discuss how to 
perform an independent measurement of the genuine temperature 
fluctuations which have to be extracted from the statistical noise.

\section{Analytical considerations}

Let us first introduce the $\Phi-$measure. One defines a single-particle 
variable $z \buildrel \rm def \over = x - \overline{x}$ with the overline 
denoting average over a single particle inclusive distribution. Here, we 
identify $x$ with $p_{\perp}$ - the particle transverse momentum. The 
event variable $Z$, which is a multiparticle analog of $z$, is defined as 
$Z \buildrel \rm def \over = \sum_{i=1}^{N}(x_i - \overline{x})$, where 
the summation runs over particles from a given event. By construction, 
$\langle Z \rangle = 0$, where $\langle ... \rangle$ represents averaging 
over events. Finally, the $\Phi-$measure is defined in the following way
\begin{equation}\label{Phi}
\Phi \buildrel \rm def \over = 
\sqrt{\langle Z^2 \rangle \over \langle N \rangle} -
\sqrt{\overline{z^2}} \;.
\end{equation}

Various fluctuations or correlations contribute to (\ref{Phi}). Our aim 
here is to compute $\Phi(p_{\perp})$ when the temperature varies from 
event to event. $ P_{(T)}(p_{\perp})$ denotes the single particle 
transverse momentum distribution in events with temperature $T$ which 
is assumed to be independent of the event's multiplicity $N$. As discussed 
in \cite{Uty01a,Wil00}, the temperature can vary within the event but we discard
such a possibility and assume that there is single temperature characterizing 
every event. We will return to this point in the concluding section. Then, 
the inclusive transverse momentum distribution reads
\begin{equation}\label{incl-dis}
P_{\rm incl}(p_{\perp}) = 
\int_0^{\infty} dT \; {\cal P}(T) \; P_{(T)}(p_{\perp}) \;,
\end{equation}
where ${\cal P}(T)$ describes the temperature fluctuations. Consequently,
\begin{equation}\label{z2}
\overline{z^2} = 
\int_0^{\infty} dT \; {\cal P}(T) \int_0^{\infty} dp_{\perp} \; 
(p_{\perp} - \overline{p_{\perp}})^2 P_{(T)}(p_{\perp}) \;,
\end{equation}
with
$$
\overline{p_{\perp}} = 
\int_0^{\infty}  dT \; {\cal P}(T) \int_0^{\infty}  dp_{\perp} \; 
p_{\perp} \; P_{(T)}(p_{\perp}) \;.
$$
The $N-$particle transverse momentum distribution in the events of 
multiplicity $N$ is assumed to be the $N-$product of $P_{(T)}(p_{\perp})$
weighed by the multiplicity and temperature distributions. Therefore, 
all inter-particle correlations, different than those due to the 
temperature variations, are entirely neglected. Then, one finds 
\begin{eqnarray}\label{Z2}
\langle Z^2 \rangle = \sum_N {\cal P}_N 
\int_0^{\infty}  dT \; {\cal P}(T)
\int_0^{\infty} dp_{\perp}^1 \;P_{(T)}(p_{\perp}^1) \; . \; . \; . 
\int_0^{\infty} dp_{\perp}^N \;P_{(T)}(p_{\perp}^N) \; 
\Big(p_{\perp}^1 +  .\; . \; . + p_{\perp}^N - N \overline{p_{\perp}}\Big)^2 \;,
\end{eqnarray}
where ${\cal P}_N$ is the multiplicity distribution. Although the 
multi-particle distribution, Eq.~(\ref{Z2}), may look as a simple 
product of the one-particle distributions, the particle distributions 
are not independent from each other due to the integration over $T$. 

In our further calculation we choose $P_{(T)}(p_{\perp})$ in the form 
suggested by the thermal model i.e.
\begin{equation}\label{pt-dis}
P_{(T)}(p_{\perp}) \sim p_{\perp}
\exp \bigg[ - {\sqrt{m^2 + p_{\perp}^2} \over T} \bigg] \;,
\end{equation}
where $m$ is the particle mass. If the transverse collective flow is taken 
into account $T$ should be understood as an {\it effective temperature} or 
simply a slope parameter controlled by the actual freeze-out temperature and 
the collective flow velocity.

In the limit $m=0$ the $p_{\perp}-$distribution (\ref{pt-dis}) acquires a
simple exponential form and one easily computes $\overline{z^2}$
and $\langle Z^2 \rangle$ given by Eq.~(\ref{z2}) and (\ref{Z2}),
respectively. Then, one gets
\begin{eqnarray*}
\overline{z^2} &=& 2\langle T^2 \rangle
+ 4 (\langle T^2\rangle - \langle T \rangle^2) \\[3mm]
{\langle Z^2 \rangle \over \langle N \rangle} &=& 2 \langle T^2 \rangle
+ 4 { \langle N^2 \rangle \over \langle N \rangle}
\big(\langle T^2\rangle - \langle T \rangle^2\big) \;,
\end{eqnarray*}
which gives 
\begin{eqnarray}\label{massless-Phi}
\Phi (p_{\perp}) = \sqrt{2} \;
{\langle N^2 \rangle - \langle N \rangle \over \langle N \rangle} \;
{\langle T^2\rangle - \langle T \rangle^2  \over  \langle T \rangle }\;,
\end{eqnarray}
when the $T-$fluctuations are sufficiently small {\it i.e.}
$\langle N \rangle  \langle T \rangle^2  \gg \langle N^2\rangle 
(\langle T^2\rangle - \langle T \rangle^2)$. For the poissonian 
multiplicity distribution the formula (\ref{massless-Phi}) 
simplifies to
\begin{eqnarray}\label{massless-poisson-Phi}
\Phi (p_{\perp}) = \sqrt{2} \; \langle N \rangle \;
{\langle T^2\rangle - \langle T \rangle^2  \over  \langle T \rangle }\;.
\end{eqnarray}

\section{Interpretation of $\Phi$}

If $T$ in Eq.~(\ref{pt-dis}) corresponds to genuine temperature,
not a slope parameter, the formula (\ref{massless-poisson-Phi}) gets
a nice interpretation due to the well-known thermodynamical relation 
\cite{Lan80} which has been discussed in the context of nuclear
collisions in \cite{Sto95,Shu98,Ste99}. Namely, 
\begin{equation}\label{relation}
{1 \over C_v} = {\langle T^2\rangle - \langle T \rangle^2 
\over \langle T \rangle^2} \;,
\end{equation}
where $C_v$ is the system's heat capacity. We note that $C_v$,
as an extensive thermodynamical parameter, is proportional 
to $\langle N_{\rm tot}\rangle$ which is the average number 
of {\it all} particles - charged and neutral - in the system at 
freeze-out, not to the average number of the {\it observed} 
particles $\langle  N \rangle$ which enters 
Eq.~ (\ref{massless-poisson-Phi}). Therefore, the formula 
(\ref{massless-poisson-Phi}) can be rewritten using the 
relation (\ref{relation}) as
\begin{equation}\label{Phi-heat}
\Phi (p_{\perp}) = \sqrt{2} \;
{\langle N \rangle  \over \langle  N_{\rm tot}\rangle}\;
{\langle T \rangle  \over c_v}\;,
\end{equation}
with $c_v = C_v/\langle N_{\rm tot} \rangle$ being the specific 
heat of hadronic matter at freeze-out. For a system of massless 
non-interacting bosons with vanishing chemical potential 
$c_v = 2 \pi^4/ 15\zeta (3) \cong 10.8$. We note that $c_v$ is 
independent of the particle's internal degrees of freedom.

A comment is in order here. It has been shown by one of us 
\cite{Mro98} that $\Phi (p_{\perp})$ vanishes in the ideal
classical gas, where the particles are exactly independent from
each other. On the other hand, the formula (\ref{Phi-heat})
tells us that $\Phi (p_{\perp})>0$ in such a gas. However, there is
no conflict between the two results. The computation from \cite{Mro98}
was performed at fixed temperature while the thermodynamical 
identity (\ref{relation}) states that there are temperature
fluctuations in any system with finite heat capacity. In fact,
these fluctuations are usually very small, because the temperature
variance is inversely proportional to the number of particles. 
Consequently, the variations of temperature are neglected in most
cases. However, $\Phi (p_{\perp})$ appears to be very sensitive to the
temperature fluctuations and the two results differ qualitatively.

In the recent paper of one of us \cite{Wil00}, the so-called non-extensive 
Tsallis statistics \cite{Tsa88} has been shown to naturally emerge when 
a system experiences temperature fluctuations. Specifically, it has been 
argued \cite{Wil00} that $1/T$ often varies according to the gamma 
distribution. Then,
\begin{equation}\label{T-dis}
{\cal P}(T) = {\alpha^{\lambda} \over \Gamma (\lambda)} 
\Big({1 \over T}\Big)^{\lambda +1} {\rm exp}\Big(-{\alpha \over T}\Big)\;,
\end{equation}
with the parameters $\lambda$ and $\alpha$ related to the moments
of $1/T$ as
$$
\Big\langle {1 \over T} \Big\rangle = {\lambda \over \alpha} 
\;,\;\;\;\;\;\;\;\;
\Big\langle {1 \over T^2} \Big\rangle - \Big\langle {1 \over T} \Big\rangle^2
= {\lambda \over \alpha^2}\;.
$$
Substituting Eqs.~(\ref{pt-dis},\ref{T-dis}) into (\ref{incl-dis}) one gets 
an inclusive distribution in the Tsallis statistics form \cite{Tsa88}
\begin{equation}\label{incl-dis-Tsallis}
P_{\rm incl}(p_{\perp}) \sim  p_{\perp}
\Big[ 1 + (q-1) \, {\sqrt{m^2 +p_{\perp}^2} \over T_0} \; \Big]^{1 \over 1-q} \;,
\end{equation}
where $T_0 \equiv \langle 1/T \rangle^{-1}$ and
$q \equiv (\lambda +1)/\lambda$ is the non-extensivity parameter \cite{Tsa88} 
related to the temperature fluctuations as \cite{Wil00}
\begin{equation}\label{q-1}
q-1 = {\big\langle {1 \over T^2} \big\rangle 
- \big\langle {1 \over T} \big\rangle^2 \over 
\big\langle {1 \over T} \big\rangle^2} 
\cong {\langle T^2 \rangle - \langle T \rangle^2 \over 
\langle T \big\rangle^2} \;.
\end{equation}
The second approximate equality holds for sufficiently small 
fluctuations. As known \cite{Tsa88}, the distribution 
(\ref{incl-dis-Tsallis}) tends to (\ref{pt-dis}) with $T=T_0$ when 
$q \rightarrow 1$. Using the relation (\ref{q-1}), the formula 
(\ref{massless-poisson-Phi}) can be rewritten in yet another form,
\begin{equation}\label{Phi-Tsallis}
\Phi (p_{\perp}) = \sqrt{2} \; \langle N \rangle \; 
\langle T \rangle \; (q-1) \;,
\end{equation}
which relates $\Phi$ to the non-extensivity parameter $q$. 

The relationship between the $\Phi-$measure and Tsallis parameter 
$q$ has been earlier considered in a different context in \cite{Alb00}. 
Namely, the authors have studied how the $q-$statistics modifies the 
usual Bose-Einstein correlations discussed in \cite{Mro98,Mro99}. Then, 
$\Phi$ has been found to decrease, not increase as in 
Eq.~(\ref{Phi-Tsallis}), with $q$.

\vspace{-0.5cm}
\begin{figure}
\centerline{\epsfig{file=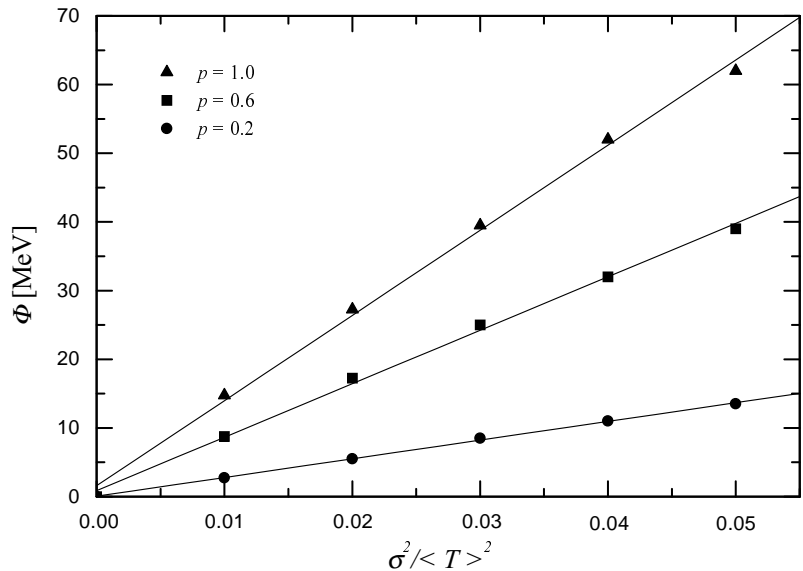,width=0.7\linewidth}}
\vspace{-0.3cm}
\caption{$\Phi (p_T)$ as a function of the temperature variance for three
values of the acceptance probability $p$. The temperature varies according
to the gaussian distribution while the multiplicity is controlled by the
lognormal one.}
\end{figure}

\section{Numerical simulation}

In this section we present the results of our Monte Carlo simulation 
of p-p collisions. The single-particle $p_{\perp}-$distribution is 
still given by Eq.~(\ref{pt-dis}). The mass equals now that of a charged 
pion because all particles in our simulation are treated as charged pions. 
We have considered two temperature distributions: the gamma-like 
form (\ref{T-dis}) and a gaussian distribution (cut off at $T<0$). The 
lognormal distribution of multiplicity of negative particles has been 
shown to fit the p-p data very well in the broad range of the collision 
energies \cite{Gaz91}. We have used the parametrization given in 
\cite{Gaz91} and assumed that the numbers of positive and negative 
pions are equal to each other in every event. The assumption is
certainly reasonable in the central rapidity domain. To check whether 
the results are sensitive to the form of the multiplicity distribution we 
have also performed a simulation with the poissonian 
distribution of negative particles. As in the case of the lognormal 
distribution, the multiplicity of charged particles has been simply 
assumed to be two times bigger than that of negative ones. The average 
charged particle multiplicity and temperature have been taken as in 
our previous paper \cite{Kor01} i.e. $\langle T \rangle = 167$ MeV 
and $\langle N \rangle = 6.56$. These values correspond to the 
proton-proton collisions at 205 GeV.

\vspace{-1.5cm}
\begin{figure}
\centerline{\epsfig{file=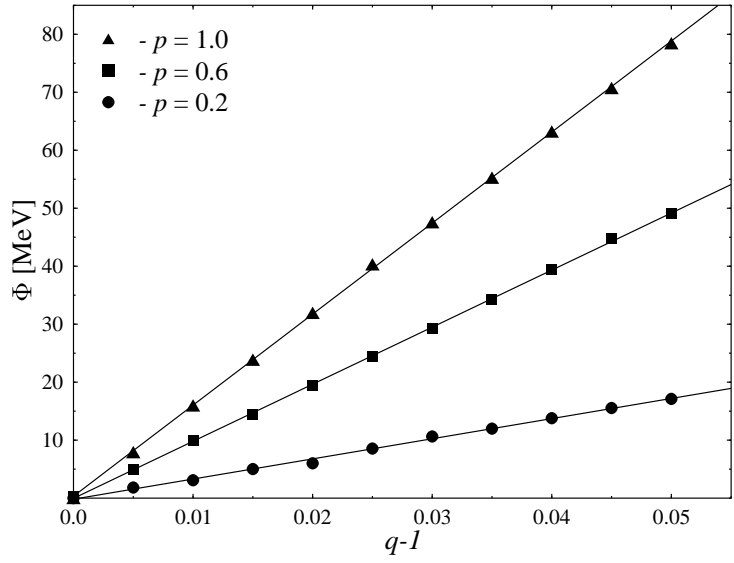,width=0.7\linewidth}}
\vspace{-0.3cm}
\caption{$\Phi (p_T)$ as a function of the temperature variance for three
values of the acceptance probability $p$. The inverse temperature varies
according to the gamma distribution while the multiplicity is controlled by
the Poisson distribution.}
\end{figure}

Due to the particle registration inefficiency and finite detector's coverage 
of the final state phase-space, only a fraction of the produced particles is 
usually observed in the experimental studies. 
Our Monte Carlo simulation takes into 
account the two effects in such a way that each generated particle - positive 
or negative pion - is registered with probability $p$ and rejected with 
$(1-p)$. The detector's acceptance usually covers a given rapidity window 
but within our model the $T-$fluctuations and $p_{\perp}-$distribution are 
rapidity independent. Therefore, there is no difference between a particle 
being lost due to the limited acceptance and one lost due to the tracking 
inefficiency.

\vspace{-0.4cm}
\begin{figure}
\centerline{\epsfig{file=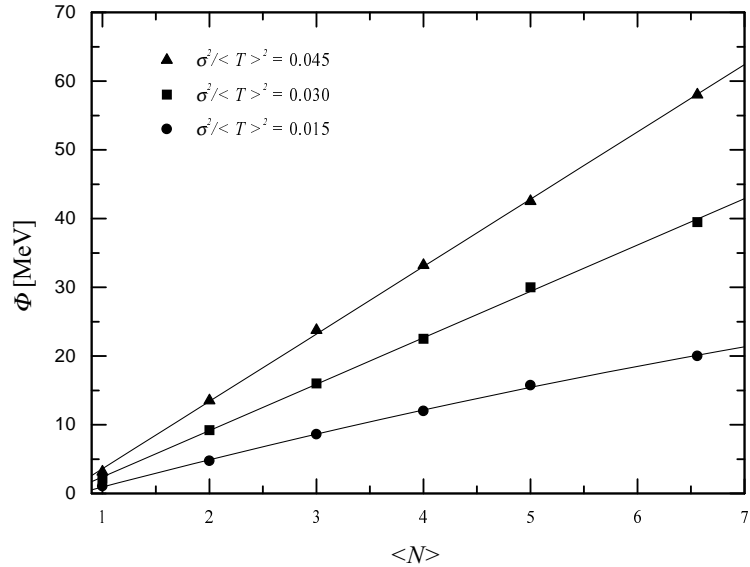,width=0.65\linewidth}}
\vspace{-0.2cm}
\caption{$\Phi (p_T)$ as a function of the average number of observed
particles for three values of the temperature variance. The temperature
varies according to the gaussian distribution while the multiplicity is
controlled by the lognormal one.}
\end{figure}

The results of our p-p simulation are shown in Figs. 1 - 4. Those in 
Figs. 1 and 3 have been found with the gaussian distribution of temperature 
and the lognormal multiplicity distribution. The results from Figs. 2 and 4 
correspond to the gamma and Poisson distributions, respectively. When 
the gamma distribution is used the $T-$variance divided by 
$\langle T \rangle^2$ is denoted by $q-1$, in agreement with Eq. (\ref{q-1}).
In the case of gaussian distribution the same quantity is written as 
$\sigma^2/\langle T \rangle^2$. One sees in Figs.~1 and 2, that 
$\Phi (p_T)$ grows linearly with the temperature variance, exactly as in Eq.~(\ref{massless-Phi}). As can also be seen, the gaussian and gamma 
distributions yield very similar results. The two observations 
mean that the effect of finite pion mass is small and that $\Phi$, 
as in the $m=0$ case, is simply a linear function of the second 
moment of $T$.

\vspace{-1.3cm}
\begin{figure}
\centerline{\epsfig{file=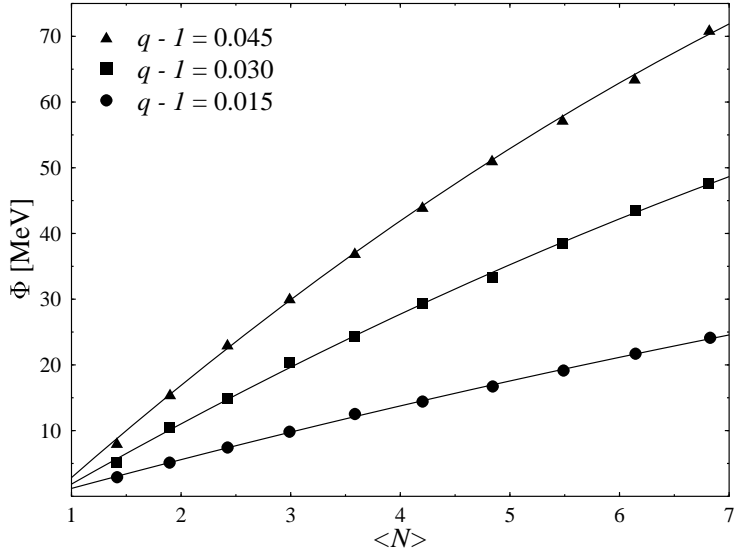,width=0.7\linewidth}}
\vspace{-0.5cm}
\caption{$\Phi (p_T)$ as a function of the average number of observed
particles for three values of the temperature variance. The inverse
temperature varies according to the gamma distribution while the
multiplicity is controlled by the Poisson distribution.}
\end{figure}

Instead of the acceptance parameter $p$ one can use the average multiplicity
of the {\it observed} particles $\langle N \rangle$ to characterize the 
acceptance. In Figs.~3 and 4 we present $\Phi (p_T)$ as a function of 
$\langle N \rangle$. The growth of $\Phi (p_T)$ with $\langle N \rangle$ 
is seen to be almost linear.

\section{Comparison with the experimental data}

The preliminary experimental value of  $\Phi (p_T) $ in proton-proton 
collisions is, as already mentioned,  $5 \pm 1$ MeV \cite{App99}. The 
measurement has been performed in the transverse momentum and 
pion rapidity intervals (0.005, 1.5) GeV and (4.0, 5.5), respectively. 
Only about 20\% of all produced particles have been observed. According 
to our simulation one needs 
$\sigma^2/\langle T\rangle^2 = q-1 \cong 0.015 \pm 0.003$, 
which corresponds to 
$\sqrt{\langle T^2 \rangle - \langle T \rangle^2}= 20 \pm 3$ 
MeV at $\langle T \rangle = 167$ MeV, to reproduce the experimental result. 

Let us now calculate the specific heat of the hadronic matter produced 
in the proton-proton interactions from the obtained value of  
$\sqrt{\langle T^2 \rangle - \langle T \rangle^2}$. 
For that we identify $\langle N_{\rm tot} \rangle$ from 
Eq.~(\ref{Phi-heat}) with the total number of pions. The pions include
those `hidden' in hadron resonances. We count each $\rho$ as two pions,
each $\omega$ as three, etc. Then, $c_v$ from (\ref{Phi-heat}) 
is the heat capacity per pion. We estimate $\langle N_{\rm tot} \rangle$
as $5 \cdot 1.5 \cdot \langle N \rangle $, where the factor  5 is due to the 
acceptance and 1.5 to include the neutral particles. Then, one finds 
from Eq.~(\ref{Phi-heat}) the heat capacity per pion $c_v = 6 \pm 2$. 
This number is significantly smaller than the previously mentioned
specific heat of massless non-interacting bosons which equals 10.8.
In fact, the discrepancy is even worse because a more realistic model
of strongly interacting matter which takes into account numerous 
resonances and, obviously, finite hadron masses gives the heat capacity 
per pion exceeding 20 \cite{Ste99}. However, the very applicability of the 
thermodynamical model to the p-p collisions is far from obvious.

As discussed in the Introduction, $\Phi(p_{\perp})$, which is measured 
in the central Pb-Pb collisions, equals $4.6 \pm 1.5$ MeV \cite{App99}. 
This value includes the short range (Bose-Einstein) correlations. 
When those correlations are excluded $\Phi(p_{\perp})=0.6 \pm 1.0$ MeV 
\cite{App99}. Since the effect of quantum statistics is not taken into 
account in our simulation, this is the latter experimental value of 
$\Phi(p_{\perp})$ which should be compared with our calculation. 
The observed average multiplicity has been 270, i.e., as in the p-p 
interactions, about 20\% of all produced charged particles \cite{App99}. 
We first identify the system freeze-out temperature with the slope 
parameter deduced from the pion transverse momentum distribution 
$\langle T \rangle \cong 180$ MeV \cite{App97}. Then, 
$\Phi(p_{\perp})=0.6 \pm 1.0$ MeV yields via 
Eq.~(\ref{massless-poisson-Phi}) 
$\sqrt{\langle T^2 \rangle - \langle T \rangle^2}= 0.5 \pm 0.4$ MeV. 
Let us stress here that our numerical simulation fully 
confirms the reliability of the analytical formula 
(\ref{massless-poisson-Phi}). The temperature is significantly 
reduced if the transverse hydrodynamic expansion is taken into 
account. The freeze-out temperature obtained by means of the 
simultaneous analysis of the single particle spectra and the
Bose-Einstein (HBT) correlations is about 120 MeV \cite{App97}.
This temperature combined with $\Phi(p_{\perp})=0.6 \pm 1.0$ MeV 
and $\langle N \rangle = 270$ gives 
$\sqrt{\langle T^2 \rangle - \langle T \rangle^2}= 0.4 \pm 0.4$ MeV. 
It has been argued in \cite{Ste99} that the transverse collective 
flow, which significantly modifies the observed temperature, does
not contribute much to the slope parameter fluctuations. Therefore,
the temperature variance is expected to be dominated by the genuine 
temperature fluctuations. 

Calculating, as in the p-p case, the heat capacity per pion   
from the temperature dispersion, one finds that 
$\langle T \rangle = 180$ MeV and 
$\sqrt{\langle T^2 \rangle - \langle T \rangle^2}= 0.5 \pm 0.4$ MeV
give $c_v = 60 \pm 100$ while $\langle T \rangle = 120$ MeV
and $\sqrt{\langle T^2 \rangle - \langle T \rangle^2}= 0.4 \pm 0.4$ MeV
correspond to $c_v = 40 \pm 70$. Unfortunately, the errors are so
large that no conclusion can be drawn.

\section{Measurement of $T-$fluctuations}

In the previous section we have shown that the temperature 
fluctuations naturally explain the p-p and Pb-Pb data. Here,
let us briefly consider how to observe independently the 
event-by-event temperature fluctuations. We discuss a 
straightforward method proposed in \cite{Ali95}. However, other 
procedures, in particular the so-called sub-event method developed 
in \cite{Vol99}, might be more efficient. The temperature variance 
can be found measuring the event's average transverse mass defined as
$$
\mu_{\perp} = {1\over N} \sum_{i=1}^Nm^i_{\perp} \;,
$$
where $N$ denotes the event's multiplicity and 
$m_{\perp}^i = \sqrt{m^2 + p_{\perp}^{i\,2}}$ is the transverse mass of 
$i-$th particle. If the single particle $p_{\perp}-$distribution is of 
the form (\ref{pt-dis}) the $m_{\perp}-$distribution reads 
\begin{equation}\label{mt-dis}
P_{(T)}(m_{\perp}) \sim m_{\perp} \: 
{\rm exp}\big[- {m_{\perp} \over T}\big] \;
\end{equation}
and $\mu_{\perp}$ is related to $T$ in the following way
$$
\mu_{\perp} = \int_m^{\infty}dm_{\perp} \: m_{\perp} \: P_{(T)}(m_{\perp}) 
= {2T^2 +2 T m + m^2 \over T+m} \;.
$$
Then, the event's temperature is expressed through $\mu_{\perp}$ as
\begin{equation}\label{T}
T = {1\over 4} \big[ \sqrt{4 m \mu_{\perp} - 4m^2 + \mu_{\perp}^2}
+ \mu_{\perp} - 2m\big] \;.
\end{equation}
Thus, measuring $\mu_{\perp}$ on the event-by-event basis one can get 
the temperature variance $V(T) = \langle T^2 \rangle - \langle T \rangle^2$. 
However, the statistical fluctuations due to the finite event multiplicity 
have to be subtracted. The point is that when the genuine temperature 
does not fluctuate at all, the observed temperature does vary because
the number of registered particles is not infinite. 

When the genuine temperature is fixed and the particles are independent 
form each other the variance of $\mu_{\perp}$ is fully determined by
the statistical fluctuations. In the events of multiplicity $N$ it
equals
$$
V_s(\mu_{\perp}) = {1 \over N}\: V_1(m_{\perp})
$$
where $ V_1(m_{\perp})$ computed with the $m_{\perp}-$distribution
(\ref{mt-dis}) is
\begin{equation}\label{V1}
V_1(m_{\perp}) = {6T^3 + 6T^2m + 3T m^2 + m^3 \over T+m}
- \Big({2T^2 +2 T m + m^2 \over T+m}\Big)^2 \;.
\end{equation}
The $T-$variance is found from the $\mu_{\perp}-$variance in the 
usual way \cite{Bra97} i.e. 
\begin{equation}\label{VsT-mu}
V_s(T) = \Big( {d T \over d \mu_{\perp}} \Big)^2 V_s(\mu_{\perp})  \;,
\end{equation}
where the derivative is computed at $ T =  \langle T \rangle$ and
$\mu_{\perp} = \langle  \mu_{\perp} \rangle$. Since $T$ is not a 
linear function of $\mu_{\perp}$ Eq.~(\ref{VsT-mu}) holds for 
sufficiently small $V_s(\mu_{\perp})$ \cite{Bra97}. This in turn 
demands that $N \gg 1$. Using Eq.~(\ref{T}), one finally finds 
the contribution of statistical fluctuations to the temperature 
variance as
\begin{equation}\label{VsT}
V_s(T) = {1 \over 16} \bigg[ 
{2 m + \langle \mu_{\perp} \rangle \over 
\sqrt{4 m \langle\mu_{\perp}\rangle - 4m^2 + \langle \mu_{\perp}\rangle^2}}
+ 1 \bigg]^2 \;
{V_1(m_{\perp}) \over N } \;,
\end{equation}
where $V_1(m_{\perp})$ is given by Eq.~(\ref{V1}) with  $T$ replaced by 
$\langle T \rangle$. We note that $V_s(T) = \langle T \rangle^2/N$
when $m=0$. The variance $V_s(T)$ should be subtracted from the
observed $T-$variance $V(T)$ to eliminate the statistical fluctuations.

\vspace{-0.5cm}
\begin{figure}
\centerline{\epsfig{file=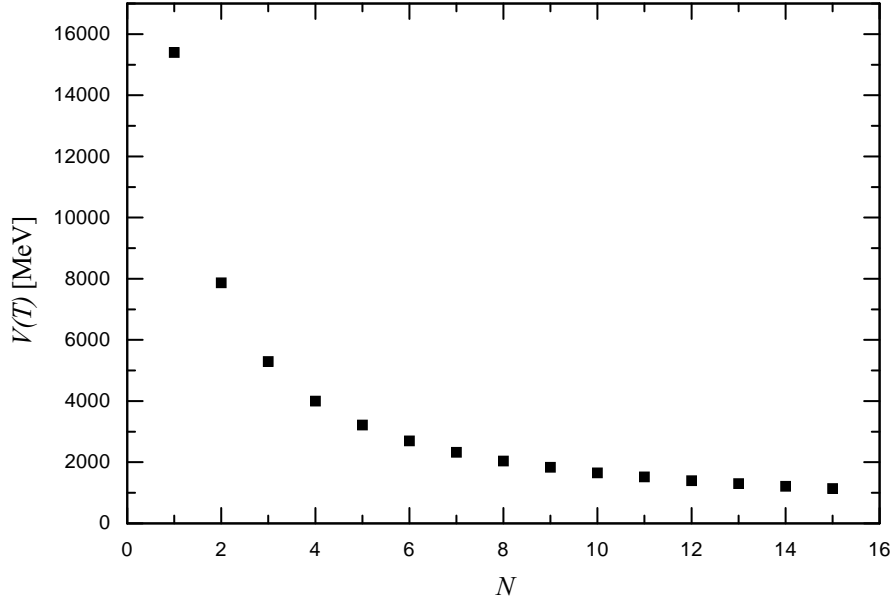,width=0.7\linewidth}}
\vspace{-0.5cm}
\caption{The observed $T-$variance as a function of multiplicity.}
\end{figure}

\vspace{-0.5cm}
\begin{figure}
\centerline{\epsfig{file=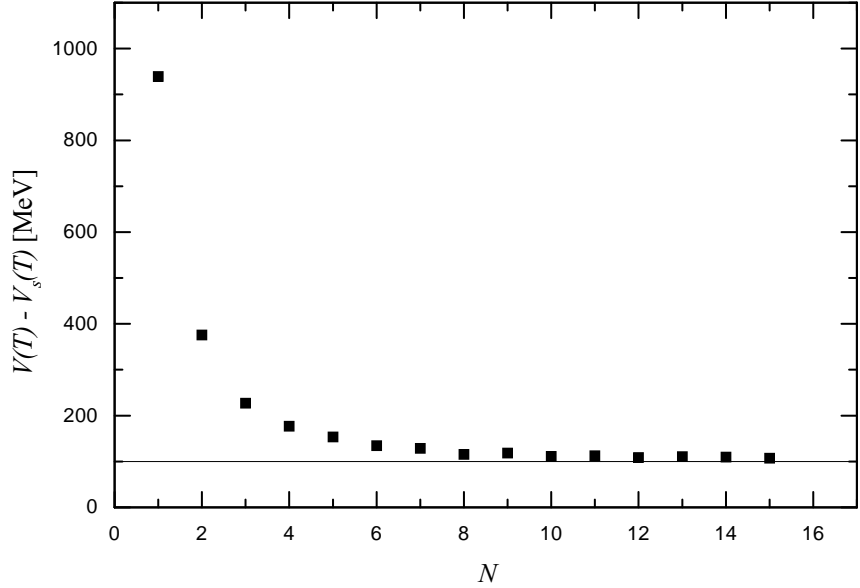,width=0.7\linewidth}}
\vspace{-0.5cm}
\caption{The $T-$variance with the subtracted statistical fluctuations 
as a function of the multiplicity.}
\end{figure}

We have performed a simple simulation to see how well the subtraction 
procedure works. For this purpose we have generated the events with fixed 
multiplicity $N$ and temperature fluctuating from event to event.
The single particle $p_{\perp}-$distribution has still been of the form 
(\ref{pt-dis}) and the temperature varied according to the gaussian 
distribution with $\langle T \rangle= 167$ MeV. The event temperature 
has been determined in the `experimental' way described above. Finally,
the temperature variance has been found and the statistical contribution
has been subtracted. In Figs. 5 and 6 we show $V(T)$ and $V(T)-V_s(T)$ 
as a function of the event multiplicity. One can see that in spite of large 
values of $V(T)$ and $V_s(T)$ the input temperature variance 
$\sigma^2 = 100 \; {\rm MeV}^2$ is well reproduced for $N$ as small
as 10. The result, although expected, is not entirely trivial. The 
expression (\ref{VsT}) assumes that the particles are {\it independent} 
from each other while in our simulation the particles are {\it correlated} 
because of the temperature fluctuations. We conclude that the measurement  
of $T-$fluctuations seems to be a feasible task even in the case of relatively 
low multiplicity collisions.

\section{Final remarks}

We have shown that $\Phi(p_{\perp})$ observed in proton-proton 
collisions \cite{App99} can be understood as an effect of temperature
fluctuations with $\sqrt{\langle T^2 \rangle - \langle T \rangle^2}= 20$ MeV. 
While the result needs to be confirmed by independent $T-$variance
measurements let us mention here  an interesting observation from \cite{Gaz01}. 
It has been found there that the transverse mass spectrum of $\pi^0$ from p-p 
collisions at $\sqrt{s} = 30$ GeV decreases as $p_{\perp}^{-P}$ over 
10 orders of  magnitude with $P=9.6$. Within the thermal model, such 
a behavior naturally appears due to the temperature fluctuations \cite{Wil00}.
Then, the exponent $P=9.6$ gives the non-extensivity parameter in 
Eq.~(\ref{incl-dis-Tsallis}) equal $q=1.10$ which translates into 
$\sqrt{\langle T^2 \rangle - \langle T \rangle^2} = 53$ MeV at 
$\langle T \rangle= 167$ MeV. The two values of the temperature dispersion 
extracted from the p-p data, 20 MeV and 53 MeV, have been found in different 
ways. One easily shows that $\Phi$ is mostly sensitive to the event-by-event 
temperature fluctuations while the inclusive distribution is shaped both
by the temperature variation from event to event and {\it within} the event. 
Therefore, the $T-$dispersion found from the inclusive spectrum is expected
to be larger than that from $\Phi$. 

$\Phi(p_{\perp})$ observed in the central Pb-Pb collisions is very 
small when the short range correlations are subtracted \cite{App99}. 
This small value corresponds to the temperature dispersion below 1 MeV.
The difference of the dispersions
found in p-p and central Pb-Pb collisions is not surprising. The hadronic 
system from nuclear interactions is not only larger - $\Phi$ as an intensive 
quantity is not directly dependent on the system size - but at freeze-out
it is expected to be much closer to the thermodynamic equilibrium. 
Consequently, the temperature fluctuations should be smaller. 

We conclude our considerations as follows. The event-by-event 
fluctuations of temperature are a possible mechanism determining
the value of $\Phi(p_{\perp})$ in p-p collisions. 
The smallness of the contribution of the long range correlations to 
$\Phi(p_{\perp})$ in the central Pb-Pb collisions is then also naturally 
explained. One needs an independent measurement of the temperature 
variance to confirm the explanation. Since the sizable fluctuations due 
to the finite statistics are, in principle, under control, such a measurement 
seems to be feasible even in the relatively low multiplicity interactions.

\begin{acknowledgements}

We are very grateful to Marek Ga\'zdzicki, Katarzyna Perl, Waldemar Retyk, 
Ewa Skrzypczak, and Grzegorz Wilk for stimulating discussions and critical
reading of the manuscript.

\end{acknowledgements}

\end{document}